\documentstyle[12pt,epsfig]{article}
\begin{document}
\baselineskip=24pt
\def\rd{{\rm d}}
\newcommand{\nn}{\nonumber}
\newcommand{\ra}{\rightarrow}
\newcommand{\be}{\begin{equation}}
\newcommand{\ee}{\end{equation}}
\newcommand{\ba}{\begin{eqnarray}}
\newcommand{\ea}{\end{eqnarray}}


\centerline{\Large \bf The Average Kinetic Energy of the Heavy
Quark in $\Lambda_{b}$} \centerline{\Large \bf in the
Bethe-Salpeter Equation Approach} \vspace{0.5cm}

\centerline{ X.-H. Guo$^{a}$\footnote{Corresponding author. email:
xhguo@bnu.edu.cn; telephone: 86-10-62209927; fax: 86-10-62231765}
and H.-K. Wu$^{b}$\footnote{email: whk302@gmail.com}}
\vspace{0.cm}

\centerline{a. Institute of Low Energy Nuclear Physics, Beijing
Normal University, Beijing 100875, China}

\centerline{b. Physics Department, Louisiana State University,
Baton Rouge, LA 70803, USA}

\vspace{0.8cm}

\centerline{\large\bf   Abstract}

\vspace{0.2cm}

In the previous paper, based on the $SU(2)_{f}\times SU(2)_{s}$
heavy quark symmetries of the QCD Lagrangian in the heavy quark
limit, the Bethe-Salpeter  equation for the heavy baryon
$\Lambda_b$ was established with the picture that $\Lambda_b$ is
composed of a heavy quark and a scalar light diquark. In the
present work, we apply this model to calculate $\mu_\pi^2$ for
$\Lambda_b$, the average kinetic energy of the heavy quark inside
$\Lambda_{b}$. This quantity is particularly interesting since it
can be measured in experiments and since it contributes to the
inclusive semileptonic decays of $\Lambda_b$ when contributions
from higher order terms in $1/M_b$ expansions are taken into
account and consequently influences the determination of the
Cabibbo-Kobayashi-Maskawa matrix elements $V_{ub}$ and $V_{cb}$.
We find that $\mu_\pi^2$ for $\Lambda_b$ is $0.25GeV^2$ $\sim$
$0.95GeV^2$, depending on the parameters in the model including
the light diquark mass and the interaction strength between the
heavy quark and the light diquark in the kernel of the BS
equation. We also find that this result is consistent with the
value of $\mu_\pi^2$ for $\Lambda_b$ which is derived from the
experimental value of $\mu_\pi^2$ for the $B$ meson with the aid
of the heavy quark effective theory.

\vspace{0.3cm}



\vspace{1cm}

\noindent{\large\bf I. Introduction}

\vspace{0.5cm}

The physics of heavy quarks has attracted intense interests in
recent years, partly because of the discovery of the flavor and
spin symmetries in QCD,  $SU(2)_{f}\times SU(2)_{s}$, in the heavy
quark limit and the establishment of the heavy quark effective
theory (HQET)~\cite{Isgur-Wise}. Compared with the research on
heavy mesons, heavy baryons have been studied less both
theoretically and experimentally. However, more and more
experimental data for heavy baryons have been and will be
obtained. This will help to test theoretical predictions for heavy
baryons. For example, the lifetime of $\Lambda_{b}$ has been
measured in several experiments~\cite{OPAL}. The measurement of
the nonleptonic decay of $\Lambda_b$,
$\Lambda_{b}\rightarrow\Lambda J/\psi$, has been done~\cite{UA1}.
There have also been the measurements of the semileptonic decays
of  $\Lambda_b$, $\Lambda_{b}\rightarrow \Lambda_c l^- \bar{\nu}_l
{\rm anything}$~\cite{ALEP} and $\Lambda_{b}\rightarrow \Lambda_c
l^- \bar{\nu}_l$~\cite{DELP}. On the other hand, since heavy
baryons are composed of three quarks instead of two, theoretical
studies for heavy baryons become more complicated. In order to
understand the hadronic structure of heavy baryons, more
theoretical and experimental studies are needed.

HQET  can simplify the physical processes involving heavy quarks.
$\Lambda_{b}$ is composed of a heavy $b$ quark and two light
quarks, $u$ and $d$.  When the heavy quark mass is very large
compared with the QCD scale $\Lambda_{QCD}$, the light degrees of
freedom (the light quark system) in a heavy baryon becomes blind
to the flavor and spin quantum numbers of the heavy quark because
of the $SU(2)_{f}\times SU(2)_{s}$ symmetries. Therefore, the
angular momentum and flavor quantum numbers of the light degrees
of freedom become good quantum numbers. Hence it is natural to
regard the heavy baryon $\Lambda_{b}$ to be composed of a heavy
quark and a light scalar diquark, $[ud]_0$, with $[ud]$ flavor
quantum number and zero spin and isospin.

The Bethe-Salpeter (BS) equation is a formally exact equation to
describe the relativistic bound state~\cite{BS, Itra, Luri}. In
the heavy quark limit the BS equation can be simplified to a great
extent and has been applied to give many theoretical results
concerning heavy mesons and  heavy baryons~\cite{Insta, guo, guo3,
guo2, guo4, XHG}. With the model for the composition of
$\Lambda_b$ which is described above the heavy baryon
$\Lambda_{b}$ is reduced from a three-body system to a two-body
system. In this picture the BS equation for $\Lambda_{b}$ was
established~\cite{guo, guo3}. A scalar confinement and a one gluon
exchange term compose the kernel of the BS equation in this model.
Furthermore, this model was generalized to the heavy baryons
$\Sigma_b^{(*)}$, $\Xi_b^{(*)}$, and $\Omega_b^{(*)}$ which are
regarded to be composed of a heavy quark and an axial-vector
diquark~\cite{guo2}.

In HQET the strong interaction of a heavy quark with four-velocity
$v$ can be described by the following Lagrangian
density~\cite{Isgur-Wise, Lag}: \be {\cal L}=\bar{h}_{v}iv\cdot
Dh_{v}+\frac{1}{2M_{Q}}\bar{h}_{v}[(iD_{\bot})^2]h_{v}
+\frac{g_{s}}{4M_{Q}}\bar{h}_{v}\sigma_{\mu\nu}G^{\mu\nu}h_{v},
\label{lag} \ee where $h_{v}$ denotes the field of the heavy
quark, $M_{Q}$ is the mass of the heavy quark,
$D^{\mu}=\partial^{\mu}-ig_sA^{\mu}$ is the covariant derivative,
$D_{\bot}=D^{\mu}- v ^{\mu}v \cdot D$, and $G^{\mu\nu}$ is the
gluon field tensor. The second operator in Eq. (\ref{lag}) is
related to the average kinetic energy of the heavy quark due to
the residual motion of the heavy quark inside the heavy hadron and
the third one corresponds to the spin energy of the heavy quark.
The kinetic energy and the spin energy of the heavy quark can be
described by the following two local matrix elements respectively:
\be \mu_\pi^2=-\frac{\langle
H_{Q}|\bar{h}_{v}(iD_\bot)^2h_{v}|H_{Q}\rangle}{2M}, \label{mupi}
\ee and \be \mu_G^2=\frac{\langle
H_{Q}|g_s\bar{h}_{v}\sigma^{\mu\nu}G_{\mu\nu}h_{v}|H_{Q}\rangle}{4M},
\label{mug} \ee where $H_Q$ ($Q=b$ or $c$) denotes a heavy baryon
containing a heavy quark $Q$ and $M$ is the mass of $H_Q$.

The parameters $\mu_\pi^2$ and $\mu_G^2$ are of particular interests
since they contribute to the inclusive semileptonic decays of heavy
hadrons when contributions from higher order terms in $1/M_Q$
expansions are taken into account and, therefore, influence the
determination of the Cabibbo-Kobayashi-Maskawa (CKM) matrix elements
$V_{ub}$ and $V_{cb}$. Therefore, it is very interesting to
calculate these nonperturbative quantities theoretically.

There have been extensive studies in literature on inclusive
semileptonic decays of bottom hadrons, $H_b \ra X e \bar{\nu}_e$,
especially since the establishment of HQET~\cite{chay, cab, ali,
jez, altare, bigi, inclusive}. These studies include corrections
to the leading order results both from perturbative QCD ($\alpha_s
(M_b)$) terms and from nonperturbative terms which are suppressed
by powers of $M_b$. It has been pointed out that there is no
$1/M_b$ corrections to the leading order result in $1/M_b$ for the
differential decay width of semileptonic decays of bottom hadrons,
$d\Gamma/dq^{2}dE_{e}$, where $q$ is total momentum of the
electron and the neutrino and $E_{e}$ is the electron
energy~\cite{chay}. Then Bigi \emph{et al.} studied $1/M_b^2$
corrections to the decay width $d\Gamma/dE_{e}$~\cite{bigi}.
Manohar and Wise analyzed extensively $1/M_b^2$ corrections to
$d\Gamma/dq^{2}dE_{e}$ for unpolarized bottom hadron $H_b$ and for
polarized $\Lambda_b$~\cite{inclusive}. In recent years,
theoretical calculations for the inclusive semileptonic decay
widths and for the moments of inclusive observables have been
carried out to order $1/M_b^3$ and $\alpha_s^2 \beta_0$
($\beta_0=11-2n_f/3$)~\cite{falk, gambino, benson, bauer,
uraltsev}. It was found that the $1/M_b^2$ corrections are
characterized by the two parameters $\mu_\pi^2$ and $\mu_G^2$,
which can be extracted from experimental data and theoretically
should been determined in a nonperturbative way.

In the case of the $B$ meson, the parameter $\mu_G^2$ can be
extracted from the data for the hyperfine splitting between $B$
and $B^*$ mesons. The other parameter $\mu_\pi^2$ has been
extracted from the experimental data for the inclusive
semileptonic $B$ meson decays $B \rightarrow X_c l \bar{\nu}$ and
$B \rightarrow X_s \gamma$~\cite{cleo, aubert1, bauer2, abdallah,
aubert2, buchmuller}. In Ref.~\cite{buchmuller}, using the
theoretical formulae provided in Refs.~\cite{gambino, benson},
Buchm\"{u}ller and Fl\"{a}cher obtained the most recent result for
$\mu_\pi^2$, $\mu_\pi^2=0.401 \pm 0.040 GeV^2$, from a combined
fit to the moments of the hadronic mass distribution and the
moments of the leptonic energy spectrum in $B \rightarrow X_c l
\bar{\nu}$ and the moments of the photon energy spectrum in $B
\rightarrow X_s \gamma$ which are measured in the $BABAR$, Belle,
CDF, CLEO, and DELPHI experiments~\cite{aubert2, babar, belle,
cdf, cleo2}. Theoretically,  $\mu_\pi^2$ has been calculated in
various phenomenological models such as QCD sum rules and the BS
equation~\cite{XHG, Eletsky, Hwang, bigi2, Gremm} and by lattice
QCD~\cite{Gimenez}. The theoretical results of $\mu_\pi^2$ for the
$B$ meson depend on models strongly. Some of them are consistent
with the experimental value, $\mu_\pi^2=0.401 \pm 0.040 GeV^2$,
while some of them, including that from the BS approach for the
$B$ meson, are not. This needs further and more careful
investigations.

Compared with the case of the $B$ meson, $\Lambda_b$ has been
studied less both experimentally and theoretically. Since
$\Lambda_{b}$ is composed of a heavy quark and a light scalar
diquark, the parameter $\mu_G^2$ is zero for $\Lambda_b$. Although
there has been no direct experimental measurement of $\mu_\pi^2$
for $\Lambda_b$, one can expect it to be measured in the future
since more and more data on $\Lambda_b$ will be collected.
Furthermore, with the aid of HQET, $\mu_\pi^2$ for $\Lambda_b$ can
be related to  $\mu_\pi^2$ for the $B$ meson~\cite{inclusive}.
Hence, one can derive the value of $\mu_\pi^2$ for $\Lambda_b$
from the experimental value of $\mu_\pi^2$ for the $B$ meson.
Therefore, it is important to give results for $\mu_\pi^2$ for
$\Lambda_b$ from theoretical calculations. The aim of the present
work is to calculate the average kinetic energy of the $b$ quark
in the heavy baryon $\Lambda_{b}$ with the BS equation model for
$\Lambda_b$~\cite{guo, guo3}. We will give the numerical result
for this parameter, discuss its dependence on the parameters in
the model, and compare our result with the value of $\mu_\pi^2$
for $\Lambda_b$ derived from the experimental value of $\mu_\pi^2$
for the $B$ meson through HQET.

The reminder of this paper is organized as the following. In
Section II we review the basic formalism for the  BS equation for
$\Lambda_b$. In Section III we give numerical solutions for the BS
wave function and then apply the BS equation to calculate
$\mu_\pi^2$ numerically. We also discuss the dependence of our
result on the parameters in the model and compare this result with
the value of $\mu_\pi^2$ for $\Lambda_b$ derived from the
experimental value of $\mu_\pi^2$ for the $B$ meson with the aid
of HQET. Finally we give a summary and discussion in Section IV.

\vspace{0.3cm}

\noindent{\large\bf II. Formalism for the BS equation for
$\Lambda_{b}$}

\vspace{0.3cm}

As discussed in Introduction, $\Lambda_{Q}$ is regarded as the
bound state of a heavy quark and a light diquark. Based on this
picture the BS wave function of $\Lambda_{Q}$ is defined as
follows: \be \chi(x_{1},x_{2},P)=\langle
0|T\psi(x_{1})\phi(x_{2})|\Lambda_{Q}\rangle, \ee where
$\psi(x_{1})$ and $\phi(x_{2})$ are field operators of the heavy
quark and the diquark, respectively, and $P$ is the momentum of
$\Lambda_{Q}$. The BS wave function in the momentum space,
$\chi_{P}(p)$, is related to $\chi(x_{1},x_{2},P)$ through the
following equation: \be
\chi(x_{1},x_{2},P)=e^{iPX}\int\frac{d^{4}p}{(2\pi)^4}\chi_{P}(p)e^{ipx},
\ee where  $p$ and $x(=x_{1}-x_{2})$ are the relative momentum and
the relative coordinate of the heavy quark and the light scalar
diquark, respectively, and $X$ is the center of mass coordinate
which is defined as $X=\lambda_{1}x_{1}+\lambda_{2}x_{2}$, where
$\lambda_{1}=\frac{M_{Q}}{M_{Q}+M_{D}}$,
$\lambda_{2}=\frac{M_{D}}{M_{D}+M_{Q}}$, with $M_{D}$ being the
mass of the diquark. The momentum of the heavy quark is
$p_{1}=\lambda_{1}P+p$ and that of the diquark is
$p_{2}=-\lambda_{2}P+p$.

The mass of the heavy baryon, $M$, satisfies the following
relation: \be M=M_{Q}+M_{D}+E_{0}+O(\frac{1}{M_{Q}}), \label{m}\ee
where $E_{0}$ is the binding energy in the leading order of
$1/M_Q$ expansion.

The BS equation in the momentum space can be written as
follows~\cite{guo}: \be
\chi_{P}(p)=S_{F}(\lambda_{1}P+p)\int\frac{d^{4}q}{(2\pi)^4}G(P,p,q)\chi_{P}(q)S_{D}(-\lambda_{2}P+p),
\label{bs}\ee where $G(P,p,q)$ is the kernel which is defined as
the sum of all the two particle irreducible diagrams with respect
to the heavy quark and the light diquark. $S_{F}$ and $S_{D}$ in
Eq. (\ref{bs}) are propagators of the heavy quark and the light
scalar diquark, respectively.

The kernel $G(P,p,q)$ includes two terms in the model: a scalar
confinement term $V_{1}$ and a one gluon exchange term
$V_{2}$~\cite{Insta, guo, guo2}, \be -iG=I \otimes
IV_{1}+v_{\mu}\otimes(p_{2}+p_{2}')^{\mu}V_{2}, \label{kernel} \ee
where $p_2$ and $p'_2$ are the momenta of the light diquark
attached to the gluon. The vertex of the gluon with the diquark
depends on the structure of the diquark. This is taken into
account by introducing a form factor $F[(p_{2}-p'_{2})^{2}]$,
which is parameterized as $F(Q^2)=\frac{\alpha_s^{\rm
eff}Q_{0}^2}{Q^2+Q_{0}^2}$~\cite{Form}, where $Q_{0}^2$ is a
parameter which freezes $F(Q^2)$ when $Q^2(=(p_{t}-q_{t})^2)$ is
very small.

It has been shown that in the leading order of $1/M_{Q}$ expansion
we only need one scalar function, $\phi_P(p)$, to describe the BS
wave function~\cite{guo}. $\phi_P(p)$ is related to $\chi_P(p)$ as
the following: \be \chi_{P}(p)=\phi_P(p) u_{\Lambda_{Q}}(v),
\label{rel}\ee where $v$ is the velocity of the heavy baryon and
$u_{\Lambda_{Q}}(v)$ is the spinor of the heavy baryon.

Define the longitudinal and transverse momenta with respect to
$v$: $p_{l}=v\cdot p-\lambda_{2}M$, $p_{t}=p-(v\cdot p)v$. Using
the covariant instantaneous approximation, $p_{l}=q_{l}$, at the
vertex of the heavy quark and the gluon, we have the BS equation
in the leading order of $1/M_{Q}$ expansion, \be
\phi_{P}(p)=\frac{-i}{(p_{l}+E_{0}+M_{D}+i\epsilon)(p_{l}^2-W_{p}^2+i\epsilon)}
\int\frac{d^{3}q_t}{(2\pi)^4} (\tilde{V_{1}
}+2p_{l}\tilde{V_{2}})\tilde{\phi}_{P}(q_t), \label{bs3}\ee where
$\tilde{\phi}_{P}(p_{t})\equiv\int(dp_{l}/2\pi)\phi_{P}(p)$ and
$\tilde{V}$ stands for $V$ in the covariant instantaneous
approximation $p_{l}=q_{l}$.

Integrating Eq. (\ref{bs3}) by $\int dp_{l}/2\pi$ and applying the
residue theorem we obtain the equation for the BS wave function,
$\tilde{\phi}_{P}(p_{t})$, \be \tilde{\phi}_{P}(p_{t})=
\frac{-1}{2W_{p}(-W_{p}+E_{0}+M_{D})}
\int\frac{d^{3}q_{t}}{(2\pi)^3} (\tilde{V_{1}
}-W_{p}2\tilde{V_{2}})\tilde{\phi}_{P}(q_{t}). \label{bs4}\ee

The kernel $\tilde{V_{1}}$ and $\tilde{V_{2}}$ have the following
expression in the case of the heavy baryon~\cite{guo, guo2}: \be
\tilde{V}_{1}=\frac{8\pi\kappa}{[(p_{t}-q_{t})^2+u^2]^2}
-(2\pi)^3\delta^3(p_{t}-q_{t})\int\frac{d^3k}{(2\pi)^3}\frac{8\pi\kappa}{(k^2+u^2)^2},
\label{v1}\ee \be \tilde{
V}_{2}=-\frac{16\pi}{3}\frac{(\alpha_{s}^{\rm
eff})^2Q_{0}^2}{[(p_{t}-q_{t})^2+u^2][(p_{t}-q_{t})^2+Q_{0}^2]},
\label{v2} \ee where $\kappa$ and $\alpha_s^{\rm eff}$ are
coupling parameters related to the scalar confinement and the one
gluon exchange diagram, respectively. The second term in Eq.
(\ref{v1}) is the counter term which removes the infra-red
divergence in the integral equation. The parameter $u$ is
introduced to avoid the infra-red divergence in numerical
calculations. The limit $u \rightarrow 0$ is taken in the end.

Substituting $\tilde{V_{1}}$ and $\tilde{V_{2}}$ into Eq.
(\ref{bs4}) we have \ba
(E_{0}+M_{D}-W_{p})\tilde{\phi}_{P}(p_{t})&=&\frac{-1}{2W_{p}}\left\{\int\frac{q_{t}^2dq_{t}}{4\pi^2}
\frac{16\pi\kappa}{(p_{t}^2+q_{t}^2+u^2)^2-4p_{t}^2q_{t}^2}\tilde{\phi}_{P}(q_{t})\right.\nonumber\\
 &&+\left.\frac{32\pi (\alpha_s^{\rm eff})^2 Q_0^2
W_{p}}{3(Q_{0}-u^2)}\int\frac{q_{t}^2dq_{t}}{4\pi^2}
\frac{1}{2p_{t}q_{t}}\left[{\rm ln}\frac{(p_{t}+q_{t})^2+u^2}{(p_{t}-q_{t})^2+u^2}\right.\right.\nonumber\\
&&-\left.\left.\ln\frac{(p_{t}+q_{t})^2+Q_{0}^2}{(p_{t}-q_{t})^2+Q_{0}^2}\right]\tilde{\phi}_{P}(q_{t})\right\}
+\frac{1}{2W_{p}}\int\frac{q_{t}^2dq_{t}}{4\pi^2}    \nonumber\\
&&\times\frac{16\pi\kappa}{(p_{t}^2+q_{t}^2+u^2)-4p_{t}^2q_{t}^2}\tilde{\phi}_{P}(p_{t}).
\label{phi}\ea

\vspace{0.5cm}

\noindent{\large\bf III. Calculation of the average kinetic energy
of the $b$ quark in $\Lambda_b$}

\vspace{0.5cm}

In this section we solve the BS equation numerically and then
apply the results to calculate the average kinetic energy of the
$b$ quark inside the heavy baryon $\Lambda_b$, $\mu^{2}_{\pi}$,
which is defined in Eq. (\ref{mupi}). The BS wave function for
$\Lambda_b$, $\tilde{\phi}_{P}(p_{t})$ in Eq. (\ref{phi}), can be
solved numerically by discretizing the integration region (0,
$\infty$) into $n$ pieces ($n$ is chosen to be sufficiently
large). We use the $n$-point Gauss quadrature rule to evaluate the
integral. Then Eq. (\ref{phi}) becomes an eigenvalue equation. The
numerical results for $\tilde{\phi}_{P}(p_{t})$ are obtained by
solving this eigenvalue equation. Eq. (\ref{phi}) is a homogeneous
equation which leaves the normalization of
$\tilde{\phi}_{P}(p_{t})$ undetermined. We use the following
normalization condition to fix the amplitude of the BS wave
function\footnote{One can also use the expression
$\mu_\pi^2=-\frac{\langle
H_{Q}|\bar{h}_{v}(iD_\bot)^2h_{v}|H_{Q}\rangle}{\langle
H_{Q}|\bar{h}_{v}h_{v}|H_{Q}\rangle}$ to calculate $\mu_\pi^2$.
This expression is independent of how the BS wave function is
normalized.}: \be \langle\Lambda_{b} |\bar{h}_{v}h_{v}|\Lambda_{b}
\rangle=2M. \ee

In the model we have several parameters, i.e. $\alpha_s^{\rm
eff}$, $\kappa$, $Q_{0}^{2}$, $M_D$ and $E_0$. The parameter
$Q_0^2$ is taken as $Q_{0}^{2}=3.2GeV^2$~\cite{guo, guo2, Form}.
The parameters $\alpha_s^{\rm eff}$ and $\kappa$ are related to
each other when we solve the eigenvalue equation with a fixed
eigenvalue~\cite{guo}. The parameter $\kappa$ varies in the region
between $0.02GeV^3$ and $0.1GeV^3$~\cite{guo}. From Eq. (\ref{m})
the parameters $M_D$ and $E_0$ are constrained by the relation
$M_{D}+E_0=M-M_{b}$ for $\Lambda_b$ in the leading order of
$1/M_b$ expansion. In our numerical calculations we use
$M_b=5.02GeV$ which leads to consistent predictions with
experiments from the BS equation in the meson case~\cite{Insta}.
Consequently we have $M_D+E_0=0.62GeV$ for $\Lambda_b$ (where we
have neglected $1/M_b$ corrections). The parameter $M_D$ can not
be determined and hence we let it vary within some reasonable
range. For $\Lambda_b$, we choose $M_D$ to be in the range
$0.65GeV$ $\sim$ $0.80GeV$. With this choice for $M_D$, the
binding energy $E_0$ is negative and varies from around $-30MeV$
to $-180MeV$.

The numerical results for $\alpha_s^{\rm eff}$ corresponding to
various values of $\kappa$ are given in Tables 1, 2, and 3. Then
the numerical results for the BS wave function depend on two
parameters, $\kappa$ and $M_D$. In Fig. 1 we show the solutions
for the BS wave function for some typical values of $\kappa$ and
$M_D$.

\begin{figure}[ht]
\begin{center}
\includegraphics*[scale=0.5,angle=-0.]{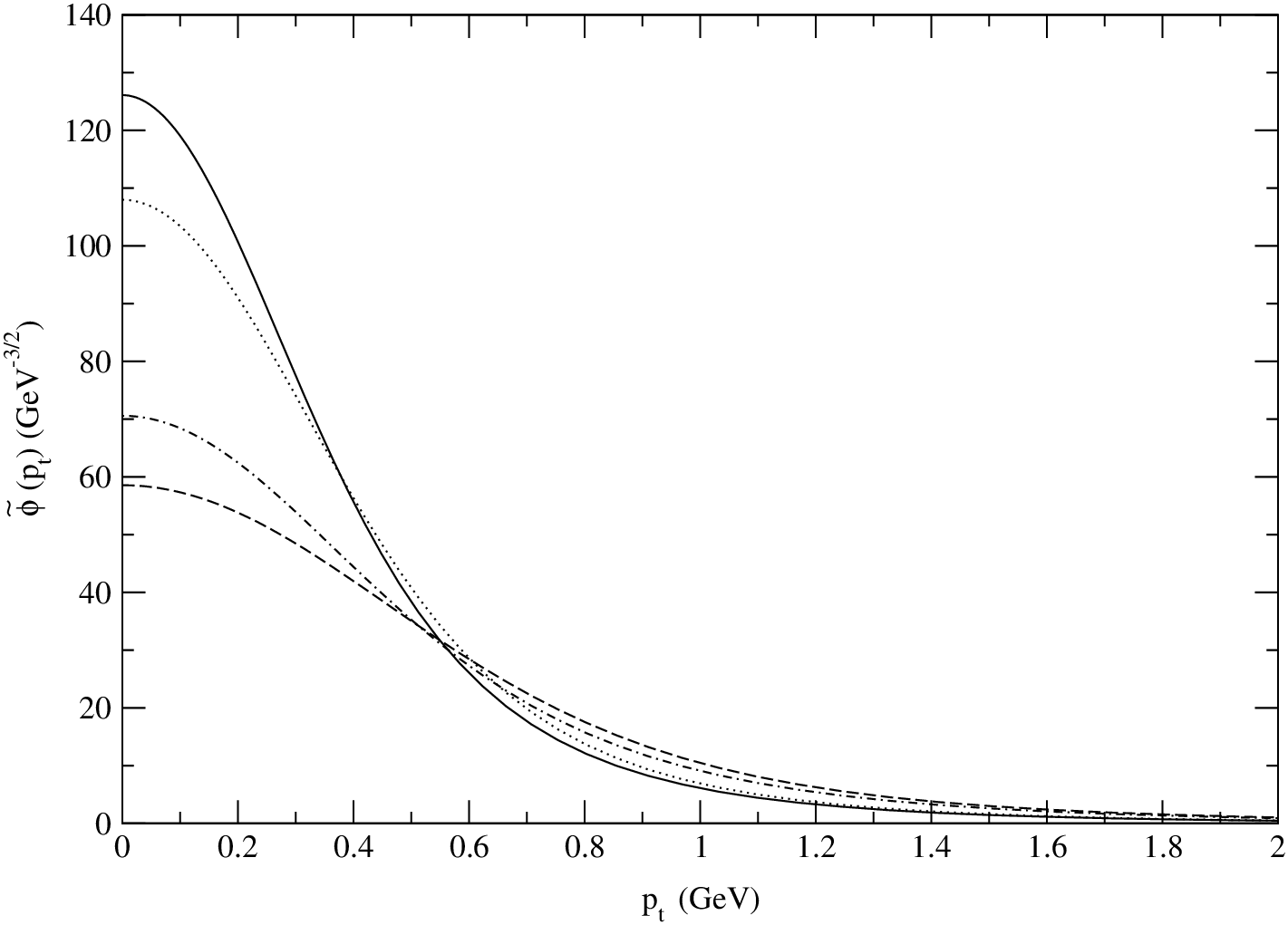}
\centerline{Fig. 1} \label{BS equation}
\end{center}
\end{figure}

Since $\mu_\pi^2$ is a Lorentz scalar~\cite{inclusive} we are free
to choose a special frame for the calculation of this parameter.
For simplicity we choose the rest frame of $\Lambda_b$ in which
Eq. (\ref{mupi}) becomes  \be
\mu_{\pi}^2=\frac{\langle\Lambda_{Q}|\bar{h}_{v}(i\vec{D})^2h_{v}|\Lambda_{Q}\rangle}{2M}.
\ee

\begin{figure}[ht]
\begin{center}
\includegraphics*[scale=0.5,angle=0.]{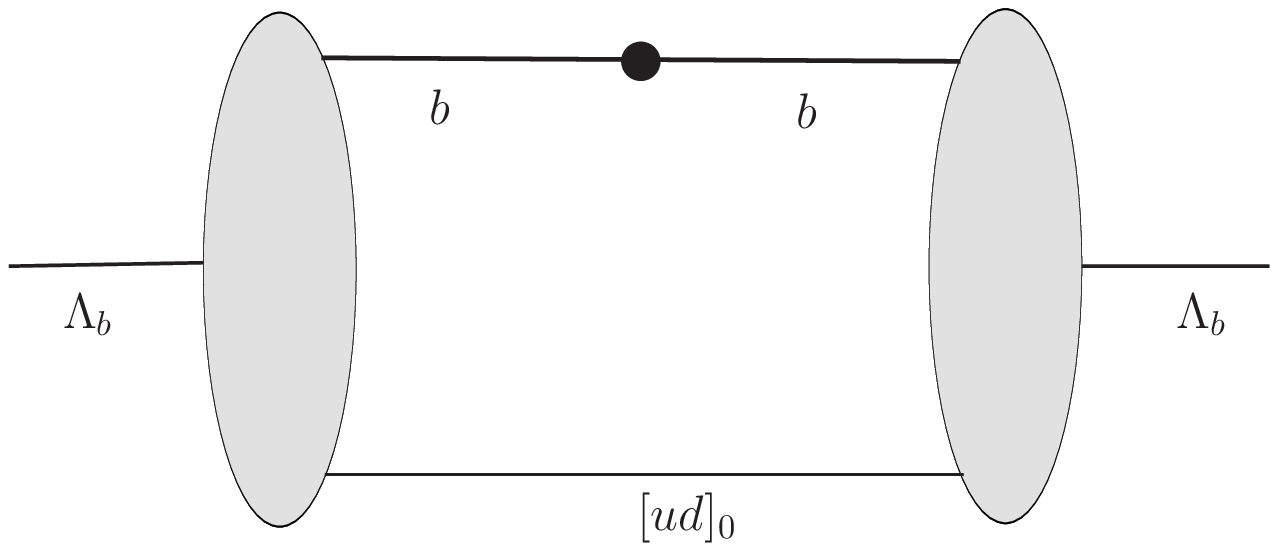}
\centerline{Fig. 2} \label{mupiplot}
\end{center}
\end{figure}

The diagram for calculating the average kinetic energy of the $b$
quark inside $\Lambda_b$ is shown in Fig. 2. Assuming the light
diquark acts as a spectator, we obtain the following expression
for $\mu_\pi^2$ which is related to the BS wave function of
$\Lambda_b$: \be
\mu_\pi^2=\frac{1}{2M}\int\frac{d^{4}p}{(2\pi)^4}\bar{\chi}_{P}(p)
 \vec{p}^2\chi_{P}(p)S^{-1}_{D}(-\lambda_{2}P+p).
 \label{mupi2}\ee

Substituting Eq. (\ref{rel}) and the relation between
$\phi_{P}(p)$ and $\tilde{\phi}_{P}(p_t)$, Eq. (\ref{bs3}), into
Eq. (\ref{mupi2}) and integrating the $p_l$ component by selecting
the proper contour we have \be \mu_{\pi}^2=\frac{1}{2M}\int
\frac{d^3p_{t}}{(2\pi)^3}
p_{t}^2(2W_{p})\tilde{\phi}^{2}_{P}(p_{t}). \label{mupi3}\ee The
three-dimensional integral in Eq. (\ref{mupi3}) can be simplified
to one-dimensional integral. This leads to \be
\mu_{\pi}^2=\frac{1}{2M}\int \frac{p_{t}^2dp_{t}}{2\pi^2}
p_{t}^2W_{p}\widetilde{\phi}^{2}_{P}(p_{t}). \label{mupi4}\ee

As shown in Fig. 1, the numerical results for the BS wave function
$\tilde{\phi}_{P}({p_{t}})$ depend on the parameters $\kappa$ (or
$\alpha_s^{\rm eff}$) and $M_D$. Therefore, the results for
$\mu_\pi^2$ also depend on these parameters. For example, taking
$M_{D}=0.7GeV$ and $\kappa=0.04GeV^3$, we get
$\mu_{\pi}^2=0.47GeV^2$.  In Tables 1, 2, and 3 we list the
numerical results for $\mu_\pi^2$ for various values of the
parameters $\kappa$ and $M_D$.

It can be seen from these tables that the value of $\mu_\pi^2$
changes from $0.25GeV^2$ to $0.95GeV^2$ in the variation ranges of
the model parameters $M_D$ and $\kappa$. The dependence of the
average kinetic energy of the $b$ quark inside $\Lambda_b$ on the
model parameters is quite strong.  Furthermore, the dependence on
$\kappa$ is stronger than that on $M_D$. For instance, for
$M_D=0.7GeV$, when $\kappa$ varies from $0.02GeV^3$ to $0.1GeV^3$,
the change of $\mu_\pi^2$ is about $0.44GeV^2$; for
$\kappa=0.04GeV^3$, when $M_D$ varies from $0.65GeV$ to $0.80GeV$,
the change of $\mu_\pi^2$ is about $0.27GeV^2$.

In principle, the parameters in the model can be determined
through the comparison between theoretical predictions and
experimental measurements about some physical processes of
$\Lambda_b$ if the data are accurate enough. In Refs.~\cite{guo,
guo3, guo4} some phenomenological predictions for $\Lambda_b$ such
as semileptonic and nonleptonic decay widths of $\Lambda_b$ are
given in the BS equation approach. Since the heavy quark mass is
not infinite in reality, in order to give more exact
phenomenological predictions $1/M_Q$ corrections to the BS
equation for $\Lambda_{Q}$ are analyzed in Ref.~\cite{guo3} based
on the assumption that $\Lambda_Q$ is composed of a heavy quark
and a scalar light diquark. Including both the $1/M_Q$ corrections
and the QCD corrections to the weak decay form
factors~\cite{neubert} the prediction for the decay rate for
$\Lambda_b \rightarrow\Lambda_c l \bar{\nu}$ can be obtained as
$2.70\sim 4.07 \times 10^{10}s^{-1}$ in the variation ranges of
$M_D$ and $\kappa$ (where $V_{cb}$ is taken as 0.042~\cite{yao}).
The uncertainty of this prediction is mostly from the uncertainty
from $\kappa$. The experimental data for the decay rate for
$\Lambda_b \rightarrow\Lambda_c l \bar{\nu}$, which is in the
range $2.3\sim 6.7 \times 10^{10}s^{-1}$~\cite{DELP}, is
consistent with the prediction. Therefore, we can not determine
the parameters in the BS equation model from this process at
present due to the large error in the experimental data. With more
and more data available in the future one can constrain the
parameters in the BS model much better. Furthermore, the
experimental data for the nonleptonic decay widths for $\Lambda_b
\rightarrow \Lambda_c$ plus a pseudoscalar or a vector meson (the
predictions for them have been given in Ref.~\cite{guo3}) can also
be used to determine the parameters in the model.

Theoretically, there have been some phenomenological calculations
for the diquark mass from the BS equation for the
diquark~\cite{yu} and from the relativistic potential model for
the diquark~\cite{ebert}, respectively. The masses for the
$[ud]_0$ diquark obtained in these two approaches depend on the
model parameters and are consistent with what are used in our BS
model.

As mentioned in Introduction, although there has been no direct
experimental measurement of $\mu_\pi^2$ for $\Lambda_b$, one can
relate this quantity to $\mu_\pi^2$ for the $B$ meson with the aid
of HQET~\cite{inclusive}. In this way, one can derive the value of
$\mu_\pi^2$ for $\Lambda_b$ from the experimental value of
$\mu_\pi^2$ for the $B$ meson. It was shown that when the masses
of heavy hadrons are expanded to order $1/M_Q$ one has the
following relation: \be \mu_{\pi}^2 (\Lambda_b) - \mu_{\pi}^2 (B)
=\frac{2M(B)M(D)}{M(B)-M(D)}\{[M(\Lambda_c)-M(D)_{\rm avg}] -
[M(\Lambda_b)-M(B)_{\rm avg}]\}, \label{relation}\ee where
$\mu_{\pi}^2 (\Lambda_b)$  ($\mu_{\pi}^2 (B)$) is $\mu_\pi^2$ for
$\Lambda_b$  ($B$), $M(B)$ ($M(D)$) is the mass of $B$ ($D$), and
$M(D)_{\rm avg}$ ($M(B)_{\rm avg}$) is defined as the spin
averaged mass of $D$ ($B$) mesons (for instance, $M(D)_{\rm avg} =
[M(D)+3M(D^*)]/4$ for $D$ mesons). The masses of $D$, $B$, and
$\Lambda_c$ have been measured accurately and the largest
uncertainty of the right hand side of Eq. (\ref{relation}) comes
from the mass of $\Lambda_b$~\cite{yao}. Using $M(\Lambda_b)=5624
\pm 9 MeV$ and the masses of $D$, $B$, and $\Lambda_c$ provided in
Ref.~\cite{yao}, the right hand side of Eq. (\ref{relation}) is
$0.025 \pm 0.052GeV^2$ where the error comes mostly from the error
of the mass of $\Lambda_b$ (the errors of the masses of $D$, $B$,
and $\Lambda_c$ contribute little). Consequently we obtain
$\mu_{\pi}^2 (\Lambda_b)$ from $\mu_{\pi}^2 (B)=0.401 \pm 0.040
GeV^2$ (which was obtained by fitting the data in the so-called
kinetic scheme~\cite{buchmuller}) as follows: \be \mu_{\pi}^2
(\Lambda_b)=0.426 \pm 0.066 GeV^2,\label{lamb}\ee where the error
includes those from both $\mu_{\pi}^2 (B)$ and the mass of
$\Lambda_b$.

\vspace{0.3cm}

Besides the uncertainty in Eq. (\ref{lamb}), the $1/M_Q^2$ terms
in the expansion for the masses of heavy hadrons may also cause
some uncertainty to $\mu_{\pi}^2 (\Lambda_b)$. Two parameters,
$\rho^3_D$ and $\rho^3$, appear in the $1/M_Q^2$ terms in the
masses of $\Lambda_b$, $\Lambda_c$, and the spin averaged masses
of $D$ and $B$ mesons\footnote{$\rho^3$ contain four terms,
$\rho^3_{\pi\pi}$, $\rho^3_{\pi G}$, $\rho^3_S$, $\rho^3_A$, while
only  $\rho^3_{\pi\pi}$ and $\rho^3_S$ contribute to the masses of
$\Lambda_b$ and $\Lambda_c$ and the spin averaged masses of $D$
and $B$ mesons~\cite{bigi2, bigi3}.}~\cite{bigi2, bigi3}. The
parameter $\rho^3_D$ has been extracted from the fit in
Ref.~\cite{buchmuller} while $\rho^3$, which is a nonlocal
correlator of the two operators $\bar{h}_v(\vec{\sigma} \cdot
\vec{D})^2 h_v$, has not been determined. $\rho^3_D$ is of order
$\bar{\Lambda}^3$ ($\bar{\Lambda}$ is defined as the difference
between the mass of a heavy hadron and the mass of the heavy quark
inside the hadron in the heavy quark limit)~\cite{buchmuller}.
$\rho^3$ is also expected to be of order $\bar{\Lambda}^3$.
Although there may be some cancellation between the parameters
$\rho^3_D$ and $\rho^3$ for the heavy baryons and those for the
heavy mesons in the mass difference $[M(\Lambda_c)-M(D)_{\rm avg}]
- [M(\Lambda_b)-M(B)_{\rm avg}]$ on the right hand side of Eq.
({\ref{relation}), we assume that the $1/M_Q^2$ terms in this mass
difference is of order $\bar{\Lambda}^3/M_Q^2$ to make a
conservative estimate on the influence of the $1/M_Q^2$ terms on
$\mu_\pi^2 (\Lambda_b)$. The $\bar{\Lambda}^3/M_c^2$ terms give
the main contribution to $\mu_\pi^2 (\Lambda_b)$ in Eq.
({\ref{relation}), which is about $0.09GeV^2$ if we take
$\bar{\Lambda}$ to be $0.6GeV$~\cite{buchmuller}.

Taking into account all the uncertainties from $\mu_\pi^2 (B)$,
the mass of $\Lambda_b$, and the $\bar{\Lambda}^3/M_c^2$ terms in
the masses of heavy hadrons, one may expect $\mu_{\pi}^2
(\Lambda_b)$ to be roughly in the range $0.27 GeV^2 \sim 0.58
GeV^2$. This is consistent with our result in the BS model, $0.25
GeV^2 \sim 0.95 GeV^2$. Conversely, one may give a rough
constraint on the ranges of the parameters in the BS model from
the range of $\mu_{\pi}^2 (\Lambda_b)$, $0.27 GeV^2 \sim 0.58
GeV^2$. For instance, when $M_D$ is $0.65GeV$, $\kappa$ is roughly
in the range $0.02 GeV^3 \sim 0.08 GeV^3$ from Table 1, while when
$M_D$ are $0.7GeV$ and $0.8GeV$, $\kappa$ are roughly in the
ranges $0.02 GeV^3 \sim 0.06 GeV^3$ and $0.02 GeV^3 \sim 0.04
GeV^3$ from Tables 2 and 3, respectively.

\vspace{0.5cm}

\noindent{\large\bf IV. Summary and Discussion}

\vspace{0.5cm}

The average kinetic energy of the $b$ quark inside $\Lambda_{b}$,
$\mu_\pi^2$, is an interesting quantity both theoretically and
experimentally. It contributes to the inclusive semileptonic
decays of $\Lambda_b$ when contributions from higher order terms
in $1/M_b$ expansions are taken into account and influences the
determination of the CKM matrix elements $V_{ub}$ and $V_{cb}$. By
comparing the experimental data with the theoretical predictions
for such decays one can extract the value of $\mu_\pi^2$.

Based on the BS equation model for the heavy baryon $\Lambda_b$,
which is regarded as composed of the heavy $b$ quark and a light
diquark, we have calculated the average kinetic energy of the $b$
quark inside $\Lambda_b$. The kernel of the BS equation consists
of a one gluon exchange term and a scalar confinement term. Since
$\mu_\pi^2$ is expressed as the overlap integral of the BS wave
function of $\Lambda_b$, we first solved out this BS wave function
numerically by transfering the integral equation for the BS wave
function into an eigenvalue equation.  We have found that the
value of $\mu_\pi^2$ varies in the region between $0.25GeV^2$ and
$0.95GeV^2$ depending on the parameters in the model. The
dependence of $\mu_\pi^2$ on the parameters in the model was
discussed in some detail. We have compared our result with the
value of $\mu_\pi^2$ for $\Lambda_b$ which is derived from the
experimental value of $\mu_\pi^2$ for the $B$ meson with the aid
of HQET and found that they are consistent. Conversely, the latter
may also be used to give a rough constraint on the parameters in
the BS model.

Compared with the meson case, heavy baryons are much more
complicated since there are three quarks in a baryon. Even though
we have simplified the bound state equation for a heavy baryon
with the diquark picture, large uncertainties are still introduced
in the BS equation for the heavy baryon. This is reflected in the
large ranges of the parameters in the model, i.e. $\kappa$ and
$M_D$. This leads to a much larger range for the phenomenological
prediction for the average kinetic energy of the $b$ quark inside
$\Lambda_b$. Fortunately much more data will be available in the
future experiments, e.g. LHCb. This provides an opportunity to
constrain the model parameters more accurately by comparing the
experimental data with the BS model predictions for the physical
processes, say semileptonic and nonleptonic decays of $\Lambda_b$.

\vspace{1cm}

\noindent{\large\bf Acknowledgements}

\vspace{0.5cm}

This work was supported in part by National Natural Science
Foundation of China (Project Number 10675022), the Key Project of
Chinese Ministry of Education (Project Number 106024) and the
Special Grants for 'Jing Shi Scholar' of Beijing Normal
University. HKW acknowledges the support from Physics Department
of Louisiana State University.

\vspace{0.5cm}

\noindent{\large\bf Figure captions}

\vspace{0.5cm}

\noindent Fig. 1 Numerical results for the BS wave function
$\tilde{\phi}_{P}({p_{t}})$. The solid (dashed) line corresponds
to $M_D=0.7GeV$ and $\kappa=0.02 (0.1)GeV^3$. The dotted
(dot-dashed) line corresponds to  $\kappa=0.04GeV^3$ and $M_D=0.65
(0.8)GeV$.

\noindent Fig. 2 The diagram for calculating the average kinetic
energy of the $b$ quark inside $\Lambda_b$. The black dot
represents the operator $\bar{h}_{v}(i\vec{D})^2h_{v}$.

\vspace{1cm}

\begin{table}[ht]
\centering
\begin{tabular}{|c|c|c|c|c|c|}
\hline
$\kappa (GeV^3)$& 0.02& 0.04 & 0.06& 0.08 & 0.10\\
\hline
$\alpha_s^{\rm eff}$&0.62& 0.67 & 0.70 & 0.72 & 0.75\\
\hline
$\mu_{\pi}^2 (GeV^2)$& 0.25&  0.39 &  0.51 &  0.62 & 0.72\\
\hline
\end{tabular}
\caption{The values of $\kappa$, $\alpha_s^{\rm eff}$, and the
corresponding $\mu_{\pi}^2$ for $M_{D}=0.65GeV$.} \label{table1}
\end{table}

\begin{table}[ht]
\centering
\begin{tabular}{|c|c|c|c|c|c|}
\hline
$\kappa (GeV^3)$&0.02& 0.04 & 0.06& 0.08 & 0.10\\
\hline
$\alpha_s^{\rm eff}$&0.67& 0.71 & 0.73 & 0.75 & 0.77\\
\hline
$\mu_{\pi}^2 (GeV^2)$& 0.34 &   0.47 &   0.58&  0.69& 0.78\\
\hline
\end{tabular}
\caption{The values of $\kappa$, $\alpha_s^{\rm eff}$, and the
corresponding $\mu_{\pi}^2$ for $M_{D}=0.7GeV$.} \label{table2}
\end{table}

\begin{table}[ht]
\centering
\begin{tabular}{|c|c|c|c|c|c|}
\hline
$\kappa (GeV^3)$&0.02& 0.04 & 0.06& 0.08 & 0.10\\
\hline
$\alpha_s^{\rm eff}$&0.76& 0.78 & 0.80 & 0.81 & 0.82\\
\hline
$\mu_{\pi}^2 (GeV^2)$& 0.55 &   0.66 &   0.76&   0.86&  0.95\\
\hline
\end{tabular}
\caption{The values of $\kappa$, $\alpha_s^{\rm eff}$, and the
corresponding $\mu_{\pi}^2$ for $M_{D}=0.8GeV$.} \label{table3}
\end{table}

\end{document}